\documentclass[conference]{IEEEtran}
\IEEEoverridecommandlockouts
\usepackage{cite}
\usepackage{amsmath,amssymb,amsfonts}
\usepackage{algorithmic}
\usepackage{graphicx}
\usepackage{textcomp}
\usepackage{xcolor}
\usepackage{multirow}
\usepackage[numbers,sort&compress]{natbib}
\usepackage{hyperref}
\hypersetup{
    colorlinks=true,
    linkcolor=blue,
    filecolor=magenta,      
    urlcolor=cyan,
}

\def\BibTeX{{\rm B\kern-.05em{\sc i\kern-.025em b}\kern-.08em
    T\kern-.1667em\lower.7ex\hbox{E}\kern-.125emX}}
\begin{document}

\title{Keystroke Biometrics in Response to \\Fake News Propagation in a Global Pandemic}

%\author{\IEEEauthorblockN{SDIM 2020 COMPSAC Workshop}
%\IEEEauthorblockA{\textit{Anonymous Submission} \\
%anonymous \\
%anonymous}

\author{\IEEEauthorblockN{\small{Aythami Morales$^1$, Alejandro Acien$^1$, Julian Fierrez$^1$, John V. Monaco$^2$, Ruben Tolosana$^1$, Ruben Vera-Rodriguez$^1$, Javier Ortega-Garcia$^1$}}
\IEEEauthorblockA{\textit{$^1$Biometrics and Data Pattern Analytics Lab, Universidad Autonoma de Madrid, Spain} \\
\textit{$^2$Naval Postgraduate School, Monterrey CA, USA}\\
\small{\{aythami.morales, alejandro.acien, julian.fierrez, ruben.tolosana, ruben.vera, javier.ortega\}@uam.es, vinnie.monaco@nps.edu}}
}

\maketitle

\begin{abstract}
This work proposes and analyzes the use of keystroke biometrics for content de-anonymization. Fake news have become a powerful tool to manipulate public opinion, especially during major events. In particular, the massive spread of fake news during the COVID-19 pandemic has forced governments and companies to fight against missinformation. In this context, the ability to link multiple accounts or profiles that spread such malicious content on the Internet while hiding in anonymity would enable proactive identification and blacklisting. Behavioral biometrics can be powerful tools in this fight. In this work, we have analyzed how the latest advances in keystroke biometric recognition can help to link behavioral typing patterns in experiments involving 100,000 users and more than 1 million typed sequences. Our proposed system is based on Recurrent Neural Networks adapted to the context of content de-anonymization. Assuming the challenge to link the typed content of a target user in a pool of candidate profiles, our results show that keystroke recognition can be used to reduce the list of candidate profiles by more than 90\%. In addition, when keystroke is combined with auxiliary data (such as location), our system achieves a Rank-1 identification performance equal to 52.6\% and 10.9\% for a background candidate list composed of 1K and 100K profiles, respectively.

\end{abstract}

%begin{IEEEkeywords}
%keystroke, de-anonymization, re-identification, fake news, COVID-19, coronavirus, biometrics.
%\end{IEEEkeywords}

\section{Introduction}

In 2020, the COVID-19 pandemic is dominating worldwide media. In an overconnected world fueled by global panic, the propagation of fake news has achieved rates never seen before.  Many of these fakes are based on ridiculous statements with scarce impact in a large percentage of the society (e.g. drinking water kills the virus or  cocaine cures the virus \cite{bbc,nypost}). However, other fake news are more sophisticated and employed to modify public opinion, propagate panic, and destabilize governments. 

The usage of fake news to manipulate public opinion has become normal in recent years, especially when major events such as elections and referendums take place. But during the COVID-19 pandemic, the spread of massive quantities of fake news has forced social media platforms to act. Companies such as Facebook or Twitter are working harder to detect and reduce the spread of fake news and bot profiles. Facebook introduced for example a “context” option that provides background information for the sources of articles in its News Feed, and Twitter has professional fact checkers to identify false content \cite{newyorktimes,forbes}. During the COVID-19 outbreak these fact checkers detected anonymous profiles publishing fake news that go directly against guidance from authoritative sources of global and local public health information, aimed to influence people into acting against recommended guidance \cite{theguardian}.

Data re-identification or de-anonymization is the practice of matching anonymous data with publicly available information, or auxiliary data, in order to discover the individual to which the data belongs to \cite{narayanan2008robust}. In the context of the fight against fake news, de-anonymization is useful to link multiple profiles belonging to the same user who is generating fake contents. Once detected, these users can be blacklisted based on their profile, MAC address, IP address, or other account data. However, these safeguards can be circumvented by creating a new account, changing device, or using a Virtual Private Network (VPN). Biometric technologies such as keystroke dynamics can be used to mitigate this circumvention. Data de-anonymization can cover any type of data, from text to audio, image, or video, being therefore a very challenging task~\cite{shu2017fake,agrawal2017multimodal,suwajanakorn2017synthesizing}. Popular examples in this line are DeepFakes, which refer to deep learning based techniques able to create fake videos by swapping the face of a person with the face of another person~\cite{2020_SurveyDeepFake_Tolosana}. In this work we focus in content that has been typed using a traditional keyboard (i.e. text). 

\begin{figure*}[t]
\centering
\includegraphics[width=\textwidth]{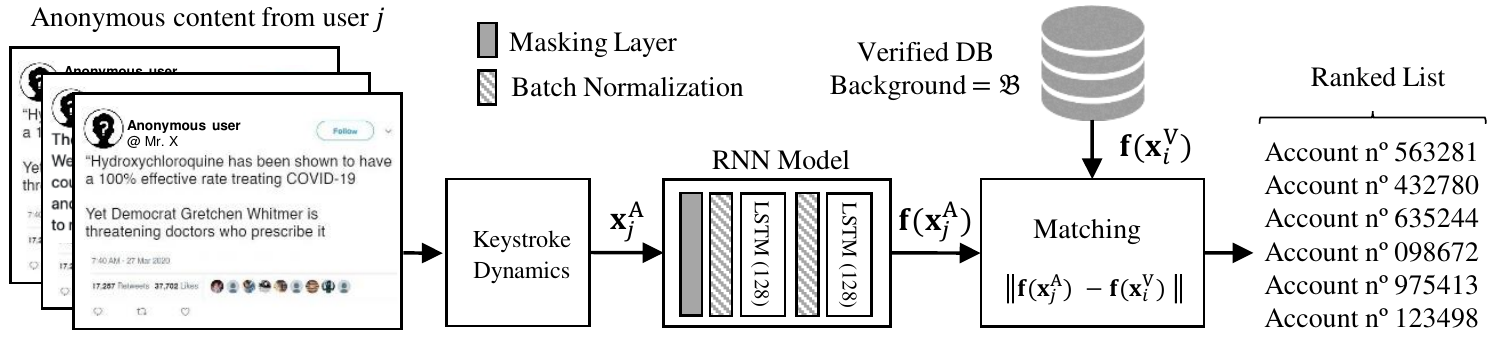}
\caption{General architecture of our proposed de-anonymization approach evaluated in this work.}
\label{Block_Diagram}
\end{figure*}

Keystroke biometric recognition enables the identification of users based on their typing behavior. During the last 15 years, the efforts of the keystroke biometrics scientific community have been mostly focused on verification scenarios with a limited number of users, typically less than several hundred. The architecture proposed in \cite{typenet2020}, with experiments conducted on over $100$,$000$ users, opened new research opportunities and challenges. The results over a user verification scenario revealed the potential of scaling up keystroke recognition. However, the suitability of this biometric trait for a large-scale identification scenario remains unexplored in the literature. 

This work presents a feasibility study of content de-anonymization based on keystroke biometrics. To the best of our knowledge, this is the first work that analyzes keystroke identification for content de-anonymization. Our results suggest the potential of keystroke identification as a tool to improve the linkability between anonymous and verified profiles.

The rest of the paper is organized as follows. Sec.~\ref{state} summarizes the state of the art in keystroke recognition. Sec.~\ref{system} defines the problem and presents the proposed system. Sec.~\ref{data} describes the experimental protocol, while in Sec.~\ref{results} we present the results. Sec. \ref{limitations} discusses the limitations and privacy concerns. Finally, Sec.~\ref{conclusions} draws the conclusions.

\section{Keystroke Biometrics: \\ From fundamentals to the State of the Art} \label{state}

Keystroke biometric systems are commonly placed into two categories: \textit{fixed-text}, where the keystroke sequence typed by the user is prefixed, such as a username or password, and \textit{free-text}, where the keystroke sequence is arbitrary, such as writing an email or transcribing a sentence with typing errors. Free-text systems must therefore consider different text content between training and testing. Biometric recognition systems can be applied for \textit{verification} or \textit{identification} task. Verification implies a 1:1 comparison to determine if the biometric sample belongs to the claimed identity. Identification implies 1:$N$ comparisons to determine the identity of the biometric sample from a pool of candidates. Biometric authentication algorithms based on keystroke dynamics for desktop and laptop keyboards have been predominantly studied for verification tasks in fixed-text scenarios, achieving accuracies higher than $95\%$ \cite{2016_IEEEAccess_KBOC_Aythami, morales2014}. Approaches based on sample alignment (e.g. Dynamic Time Warping) \cite{2016_IEEEAccess_KBOC_Aythami}, Manhattan distances \cite{Vinnie1}, digraphs \cite{Bergadano}, and statistical models (e.g. Hidden Markov Models) \cite{Ali} have achieved the best results in fixed-text verification.

However, the performances of free-text algorithms are generally far from those reached in the fixed-text scenario, where the complexity and variability of the text entry contribute to intra-subject variations in behavior, challenging the ability to recognize users \cite{Sim}. Monrose and Rubin \cite{Monrose} proposed a free-text keystroke algorithm based on user profiling by using the mean latency and standard deviation of digraphs and computing the Euclidean distance between each test sample and the reference profile. Their results worsened from $90\%$ to $23\%$ of correct classification rates when they changed both users’ profiles and test samples from fixed-text to free-text. Gunetti and Picardi \cite{Gunetti} extended the previous algorithm to n-graphs. They calculated the duration of n-graphs common between training and testing and defined a distance function based on the duration and order. Their results of $7.33\%$ classification error outperformed the previous state of the art. %Nevertheless, their algorithm needs long keystroke sequences (between $700$ and $900$ keystrokes) and many keystroke sequences (up to $14$) to build the user´s profile, which limits the usability of that approach. Murphy \etal~\cite{Murphy} more recently collected a very large free-text keystroke dataset ($\sim$$2.9$M keystrokes) and applied the Gunetti and Picardi algorithm achieving $10.36\%$ classification error using sequences of $1$,$000$ keystrokes and $10$ genuine sequences to authenticate users.

Recently, some algorithms based on statistical models have been shown to work very well with free-text, like the POHMM (Partially Observable Hidden Markov Model) \cite{Monaco}. %This algorithm is an extension of the traditional Hidden Markov Models (HMMs), but with the difference that each hidden state is conditioned on an independent Markov chain. This algorithm is motivated by the idea that keystroke timings depend both on past events and the particular key that was pressed. 
Performance achieved using that approach in free-text is close to fixed-text, but requires several hundred keystrokes and has only been evaluated with a database containing less than 100 users. The latest advances in deep learning and the availability of large scale databases has boosted the performance of free-text keystroke recognition biometrics only very recently. In \cite{typenet2020}, a Deep Recurrent Neural Network architecture was presented with experiments over a database with $168$,$000$ users and $136$M keystrokes. Results obtained within a free-text verification scenario achieved error rates under $5\%$. Nevertheless, the performance of these algorithms for large scale identification scenarios remains unknown. This is one of the major contributions of this work.

\begin{figure*}[t!]
\centering
\includegraphics[width=\textwidth]{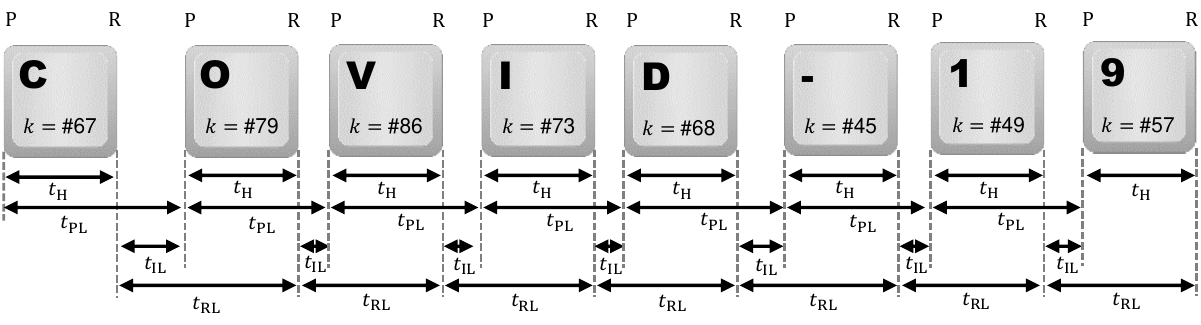}
\caption{Example of the 37 temporal features extracted from the term "COVID-19": $8$ $\times$ Hold Time ($t_{\textrm{H}}$) + $7$ $\times$ Inter-key Latency ($t_{\textrm{IL}}$), $7$ $\times$ Press Latency ($t_{\textrm{PL}}$), $7$ $\times$ Release Latency ($t_{\textrm{RL}}$), $8$ $\times$ key codes ($k$). P = key Press event; R = key Release event.}
\label{features_fig}
\end{figure*}

\section{Problem Statement and System Description} \label{system}

\subsection{Problem Statement}

Anonymous content is re-identified when multiple anonymous profiles are linked or associated to a verified profile. In our experiments, we assume that fake content was typed by an anonymous user who authored other verified content published on the same or different platform. A \textit{verified content} is defined in this work as a content associated to a real identity (e.g. personal social media account or digital profile certified by a third party). An \textit{anonymous content} is defined as a content published by one user who has not revealed his/her real identity (i.e. this content is usually associated with an alias or pseudonym).

In this work, we assume that timing sequences of the keyboard were captured when typing. This can occur when a user types content directly into a webpage. No special permissions are required to record the timestamps of keyboard input events generated on all major web browsers (e.g., Chrome, Firefox, Safari), and only the Tor Browser attempts to obfuscate typing behavior by lowering timestamp resolution \cite{torbrowser}. De-anonymization is achieved by comparing the typing characteristics of the anonymous and the verified contents using the biometric patterns associated to the keystroke dynamics derived from these timing sequences. Fig. \ref{Block_Diagram} presents the architecture of our proposed approach. The Anonymous typed content is first characterized according to the keystroke dynamics $\textbf{x}^\textrm{A}$ (A for Anonymous) associated to the sequences of time events $t$ and keycodes $k$. A Recurrent Neural Network is used to project the timing and keycode sequences into a feature space trained for keystroke verification. The generated feature vector $\textbf{f}(\textbf{x}^{\textrm{A}})$ is characterized by the individual typing behavior of the anonymous subject. The feature vector $\textbf{f}(\textbf{x}^{\textrm{A}})$ is then matched with each of the $i$ feature vectors $\textbf{f}(\textbf{x}_i^{\textrm{V}})$ of a Verified content database $\mathfrak{B}$ (the ``background''), composed of $N=\#\mathfrak{B}$ profiles. The result of the matching process is a ranked list with the $N$ profiles ordered by similarity to the anonymous subject. % (i.e. firs rank corresponds to the most likely profile to be owned by the anonymous user, second rank to the second most likely,...).

\subsection{Pre-processing and Keystroke Dynamics} \label{features}
The raw data captured in each keystroke sequence is composed of a three dimensional time series including: the keycodes, key press timestamps (corresponding to keydown events), and key release timestamps (corresponding to keyup events). In our experiments, timestamps were in UTC format with millisecond resolution (captured in a web browser), and the keycodes were integers between $0$ and $255$ according to the ASCII code.

%AM: Thanks! I have accepted your suggestions, much better now
From this raw data, we extract $4$ temporal features popular in keystroke recognition (see Fig. \ref{features_fig} for details): (i) Hold Time ($t_{\textrm{H}}$): the elapsed time between press and release events; (ii) Inter-key Latency ($t_{\textrm{IL}}$): the elapsed time between releasing a key and pressing the next key; (iii) Press Latency ($t_{\textrm{PL}}$): the elapsed time between two consecutive press events; and Release Latency ($t_{\textrm{RL}}$): the elapsed time between two consecutive release events. These $4$ features are commonly used in both fixed-text and free-text keystroke systems \cite{Alsultan}. Finally, we include the keycodes as an additional feature.

Let $L$ be the length of the keystroke sequence. The keycode and Hold Time features are calculated for each of the $L$ keys in the sequence, and latency features between consecutive keys ($t_{\textrm{IL}}$, $t_{\textrm{PL}}$, and $t_{\textrm{RL}}$) are calculated for the $L-1$ consecutive key pairs. This produces a time series with shape $L \times 2 + (L-1) \times 3$. All feature values are normalized before being provided as input to the model. Normalization is important so that the activation values of neurons in the input layer of the network do not saturate (i.e. all close to $1$). The keycodes are normalized between $0$ and $1$ by dividing each keycode by $255$, and the $4$ timing features are converted to seconds. This scales most timing features between $0$ and $1$ as the average typing rate over the entire dataset is $5.1$ $\pm$ $2.1$ keys per second. Only latency features that occur either during very slow typing or long pauses exceed a value of $1$.

\subsection{Recurrent Neural Network Architecture} \label{architecture}

We employ the Recurrent Neural Network (RNN) model proposed in \cite{typenet2020}. The model is composed of two Long Short-Term Memory (LSTM) layers of $128$ units. Between the LSTM layers there are batch normalization and dropout layers ($0.5$ drop rate) to avoid overfitting. Additionally, each LSTM layer has a $0.2$ recurrent dropout rate. The network was trained with more than $1$M keystroke sequences (over $50$M keystrokes) from $68$,$000$ different users (see Sec. \ref{protocol} for details).

The RNN was trained using a Siamese setup involving two inputs: two keystroke sequences from either the same or different users. During the training phase, the model learns the projections necessary to discriminate whether two keystroke sequences belong to the same user or not. The model acts as a feature extractor and outputs an embedding vector that contains the discriminating features (see \cite{typenet2020} for details).

One constraint when training a RNN using standard backpropagation through time applied to a batch of sequences is that the number of elements in the time dimension (i.e. number of keystrokes) must be the same for all sequences. We fix the size of the time dimension to $M$. In order to train the model with sequences of different lengths $L$ within a single batch, we truncate the end of the input sequence when $L>M$ and zero pad at the end when $L<M$, in both cases to the fixed size $M$. Error gradients are not computed for zeroed elements, which do not contribute to the loss function in the iterative learning due to the Masking layer indicated in Fig.~\ref{Block_Diagram}.

Finally, the output of the RNN model $\textbf{f}($\textbf{x}$)$ is an array of size $1 \times 128$ that we consider later as an embedding feature vector to identify anonymous content based on Euclidean distance.

\begin{figure*}[t!]
\centering
\includegraphics[width=1\textwidth]{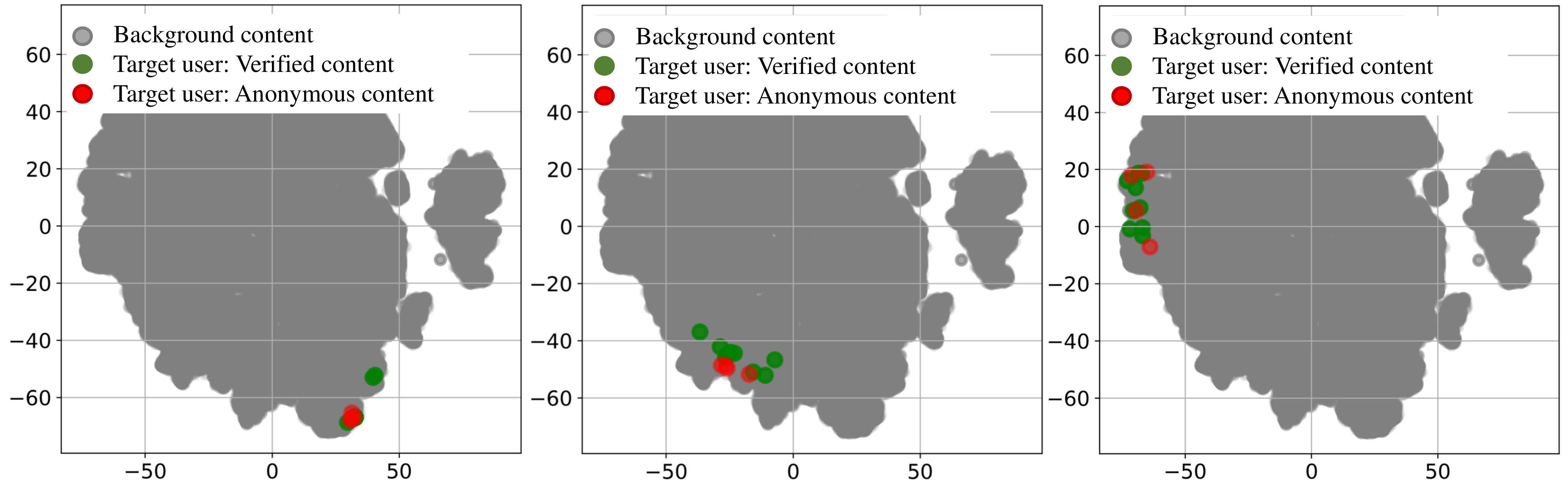}
\caption{t-SNE projections of $30$,$000$ embedding vectors from $2$,$000$ users. Each projection is obtained from the embedding generated by the Recurrent Neural Network. Each sub-figure contains the projections of the Anonymous and Verified keystroke sequences of a different target user. (Color image)}
\label{tsne}
\end{figure*}

\section{Dataset and Experimental Protocol} \label{data}

\label{experimental_protocol}
\subsection{Dataset}

\label{aalto}
All experiments were conducted with the Aalto University Dataset \cite{Dhakal} that comprises keystroke data collected from $168$,$000$ participants during a three-month time span. The acquisition task required subjects to memorize English sentences and then type them as quickly and accurately as they could. The English sentences were selected randomly from a set of $1$,$525$ examples taken from the Enron mobile email and Gigaword newswire corpora. The example sentences contained a minimum of $3$ words and a maximum of $70$ characters. Note that the sentences typed by the participants could contain more than $70$ characters as each participant could forget or add new characters when typing.

For the data acquisition, the authors launched an online application that recorded the keystroke data from participants who visited their website and agreed to complete the acquisition task (i.e. the data was collected in an uncontrolled environment). Press (keydown) and release (keyup) event timings were recorded in the browser with millisecond resolution using the JavaScript function \texttt{Date.now}. All participants in the database completed $15$ sessions (i.e. one sentence for each session) on either a physical desktop or laptop keyboard. The authors also reported demographic statistics: $72$\% of the participants took a typing course, $218$ countries were involved, and $85$\% of the participants had English as native language.

\subsection{Experimental Protocol} \label{protocol}

The RNN was trained using the first $68$,$000$ users in the dataset according to the method proposed in \cite{typenet2020}. The size of the time dimension $M$ was fixed to $M=50$, which is short enough to consider small sentences. The remaining $100$,$000$ users will be employed only to perform the evaluation of the de-anonymization system, so there is no data overlap between the two groups of users. Note that the model is unique for all the $100$,$000$ users in the evaluation set, and does not require specific training when a new target user is added.  

The $15$ sequences from the $100$,$000$ users in the database were divided into two groups that simulate a de-anonymization scenario: Verified ($10$ sequences) and Anonymous ($5$ sequences). We evaluated the de-anonymization accuracy by comparing the Anonymous set of samples $\textbf{x}_{j,l}^{\textrm{A}}$, with $l=1,...,5$ belonging to the user $j$ against the Background Verified set $\textbf{x}_{i,g}^{\textrm{V}}$, with $g=1,...,10$  belonging to all $100$,$000$ users. The distance was computed by averaging the Euclidean distances $||\cdot||$ between each Verified embedding vector $\textbf{f}(\textbf{x}_{i,g}^{\textrm{V}})$ and each Anonymous embedding vector $\textbf{f}(\textbf{x}_{j,l}^{\textrm{A}})$ as follows:
\begin{equation}
\label{score}
     d_{i,j}= \frac{1}{10 \times 5}\sum_{g=1}^{10}\sum_{l=1}^{5} ||\textbf{f}(\textbf{x}_{i,g}^{\textrm{V}})-\textbf{f}(\textbf{x}_{j,l}^{\textrm{A}})||
\end{equation}

We then re-identify an anonymous profile (i.e. Anonymous subject $j=J$ is the same Verified person $i=I$) as follows:

\begin{equation}
\label{mindistance}
     I = \arg\min_i d_{i,J}
\end{equation}

The results reported in the next section are computed in terms of Cumulative Match Curve (CMC), which is a measure of $1$:$N$ identification system performance. The curves are calculated for each user and then averaged over all $100$,$000$ users. A Rank-$1$ means that $d_{i,J}<d_{I,J}$ for any $i \neq I$, while a Rank-$n$ means that instead of selecting a single Verified profile, we select $n$ of them starting with $i=I$ by increasing distance $d_{i,J}$. In forensic scenarios, it is traditional to use Rank-20, Rank-50, or Rank-100 in order to generate a short list of potential candidates that are finally identified manually using a bag of evidence.

\section{Experimental Results} \label{results}

\subsection{Discriminatory Potential of Keystroke Biometrics}

To ascertain the potential of the feature vectors generated by the keystroke model, we applied the popular data visualization algorithm t-SNE over the dataset. t-SNE is an algorithm to visualize high-dimensional data. This algorithm minimizes the Kullback-Leibler divergence between the joint probabilities of the low-dimensional embedding and the high-dimensional data. Fig. \ref{tsne} shows the projection of the keystroke embedding of three target users into a 2D space generated by the t-SNE algorithm. t-SNE projection is an unsupervised algorithm but for interpretation purposes we have colored three groups:

\begin{itemize}
    \item Background projections including embeddings of $30$,$000$ keystroke sequences from $2$,$000$ random users. These sequences serve to visualize the boundaries and shape of the feature space represented by t-SNE.
    \item Validated sequences ($10$ per user) typed by $3$ target users (each one in a different plot). These sequences model the target user's typing behavior. 
    \item Anonymous sequences ($5$ per user) typed by the $3$ target users. These sequences serve to re-identify the target users. 
\end{itemize}
 
As we can see, the projections of the Validated and Anonymous keystroke embeddings from the target users are represented in close regions of the t-SNE space. Note that the t-SNE projection is not trained using the labels (i.e. identity) associated to each keystroke embedding. As a result, we observe that the personal typing patterns from the target users as represented by our deep network are discriminative enough in this large background set of $30$,$000$ different sequences. These results suggest the discriminatory information available in this biometric trait.    

\begin{figure}[t!]
\centering
\includegraphics[width=0.95\columnwidth]{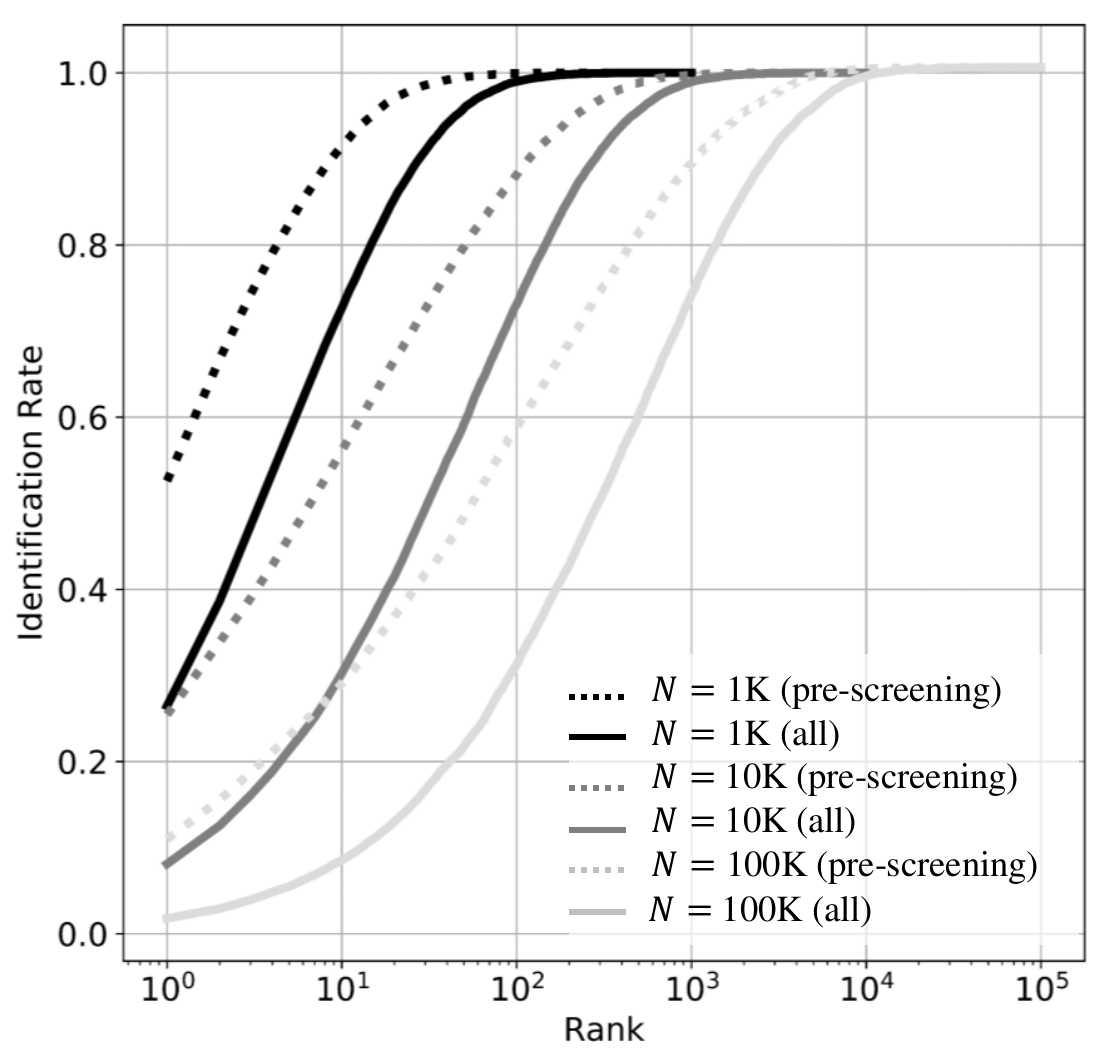}
\caption{CMC curves for different background sizes $N=\#\mathfrak{B}$. Pre-screening is based on the assumption that the location of the typist is available and can be used to reduce the number of candidates in the background set.}
\label{rank}
\end{figure}

\subsection{De-anonymization Accuracy}

As a $1$:$N$ problem, the identification accuracy varies depending on the size of the background set $\mathfrak{B}$. In our experiments, the size of the background is equal to the number of verified profiles available for the comparison.  Fig. \ref{rank} and Table \ref{table_acc} present the de-anonymization performance for different background sizes. The results vary depending on the size of the background, with Rank-1 identification accuracy varying from $1.8\%$ for the largest background ($100$K) to $26.4\%$ for the smallest one ($1$K). These accuracies can be considered low for verification scenarios associated to user authentication. But in the context of profile de-anonymization, a $1.8\%$ accuracy among $100$,$000$ profiles means that $1$,$800$ profiles can be automatically de-anonymized using only the keystroke patterns of the owner. Rank-1 identification rate reveals the ability to unequivocally identify the anonymous target profile among all the verified profiles in the background set. Additionally, Rank-$n$ represent the achievable accuracy if we consider a ranked list of $n$ profiles from which the de-anonymization is then manually or automatically conducted based on additional evidence \cite{2018_INFFUS_MCSreview2_Fierrez}. In general, the results suggest that keystroke de-anonymization enables a $90\%$ size reduction of the candidate list while maintaining $100\%$ accuracy (see the CMC curves in Fig. \ref{rank}). 

The number of background profiles can be further reduced if auxiliary data is available to realize a pre-screening of the initial list of verified profiles (e.g. country, language). The Aalto University Dataset contains auxiliary data including age, country, gender, type of keyboard, and others. Fig. \ref{rank} and Table \ref{table_acc} show also user identification accuracy over the entire background dataset with a pre-screening by country (i.e., contents generated in a country different to the country of the target user are removed from the background set). The results show that pre-screening based on a unique attribute is enough to largely improve the identification rate: Rank-1 identification with pre-secreening ranges between $10.9\%$ to $52.6\%$, while the Rank-100 ranges between $58.8\%$ to $99.9\%$. These results demonstrate the potential of keystroke dynamics for de-anonymization when auxiliary information is available.

\begin{table}[t]
\normalsize
  \begin{center}
    \caption{Identification accuracy (Rank-$n$ in \%) for different background sizes $N$. In brackets accuracy with pre-screening of the background dataset based on the location of the typist.}\smallskip
    \label{table_acc}
    \begin{tabular}{|l|c|c|c|} 
     % \toprule
      \hline
      \multirow{2}{*}{\textbf{Rank}} & \multicolumn{3}{c|}{\textbf{Background Size $N$}} \\
      \cline{2-4}
      & $N$ = 1K & $N$ = 10K & $N$ = 100K \\
     \hline
      %\midrule % <-- Midrule here
      %\Tspace
      Rank-1 & $26.4$ $(52.6)$ & $8.1$ $(25.4)$ & $1.8$ $(10.9)$ \\ % <-- Content of first column omitted.
      \hline
      %\midrule % <-- Midrule here
      %\Tspace
      Rank-50 & $95.9$ $(99.5)$ & $59.2$ $(79.7)$  & $21.8$ $(48.4)$ \\
      \hline
      Rank-100 & $98.9$ $(99.9)$ & $73.1$ $(88.1)$  & $31.3$ $(58.8)$ \\
      \hline
      Rank-1000 & $100$ $(100)$ & $98.9$ $(99.7)$  & $74.2$ $(89.4)$ \\
      \hline
      %\midrule % <-- Midrule here
      %\Tspace
      Rank-5000 & $-$ & $99.9$ $(99.9)$  & $96.1$ $(99.6)$ \\
      \hline
      
      %Ours (VGG-Face) & 10.03(21\%) & 1.98 (43\%) & 1.69 (61\%)\\ % <-- Content of first column omitted.
      %\bottomrule % <-- Bottomrule here
    \end{tabular}
  \end{center}
\end{table}

\section{Limitations and Privacy Aspects}
\label{limitations}

The results presented in this work are encouraging. However, there are still some limitations regarding the application of this technology in the fight against fake news. The first one is that content must be typed. Spread of news by retweet or similar sharing mechanisms which do not require use of the keyboard, are not detectable by this technology. Second, bots are commonly employed for the propagation of fake content. It is not clear how the method proposed in this work would perform for synthetic behavior emulated by bots \cite{becaptcha}. Third, the identification performance decays for a large number of background profiles. Therefore, pre-screening is recommendable to reduce the candidate list.   

On the other hand, biometric data is considered sensitive data in a number of regulations (e.g. paragraph 71, EU GDPR). Keystroke dynamics, as biometric data, must be processed according to appropriate technical and organizational methodologies. The proposed de-anonymization based on keystroke behaviors is a powerful tool that can help in the fight against missinformation. But at the same time, the missuses of this technology arise important concerns related to data protection and user privacy. In addition to the identification accuracy studied in the paper, a concrete application of the ideas developed here should also consider and evaluate a secure storage of the biometric templates, and other modules for privacy preservation \cite{bringer2013}. The balance is delicate in this case, as we should aim to de-anonymize problematic subjects while preserving the privacy rights and freedom of speech of the overall population at the same time. Careful consideration of such security and privacy aspects is out of the scope of the present paper and can be investigated elsewhere \cite{Kindt2013,Campisi2013}.

%[http://biometrics.eps.uam.es/fierrez/showbibtex.php?id=2017_Access_HEmultiDTW_Marta]

\section{Conclusions} \label{conclusions}

This work proposes keystroke biometric recognition for typed content de-anonymization. The fight against missinformation requires new tools, and the COVID-19 pandemic has showed the necessity to develop new technologies and policies to reduce the spread of fake content. Keystroke recognition can be used as a tool to link multiple profiles belonging to the same typist based on his typing behavior. We have evaluated a system based on Recurrent Neural Networks in experiments involving $100$,$000$ users and more than $1$M keystroke sequences. Our results suggest the potential of this technology to link multiple texts typed by the same user by leveraging personal typist patterns. The performance achieved varies depending on the number of background profiles, with Rank-1 identification accuracy ranging from $10.9\%$ to $52.6\%$ and Rank-50 from $48.4\%$ to $99.5$ when auxiliary information is available. 

\section*{Acknowledgments}

This work has been supported by projects: PRIMA (MSCA-ITN-2019-860315), TRESPASS (MSCA-ITN-2019-860813), BIBECA (RTI2018-101248-B-I00 MINECO), BioGuard (Ayudas Fundación BBVA a Equipos de Investigación Científica 2017). A. Acien and R. Tolosana are supported by a FPI and postdoc fellowship from the Spanish MINECO.

{\small
\bibliographystyle{ieee}
\bibliography{submission_example.bib}
}

\end{document}